\documentclass[showpacs,amsmath,amssymb,superscriptaddress,aps,a4paper,prl,twocolumn]{revtex4}
\usepackage{graphicx}
\usepackage{dcolumn}% Align table columns on decimal point
\usepackage{bm}% bold math
\usepackage{textcomp}
\usepackage{color}

\begin{document}

\title{Observation of magneto-phonon resonance of Dirac fermions in graphite}

\author{Jun Yan}
\affiliation{Department of Physics and Department of Applied Physics
and Applied Mathematics, Columbia University New York, USA}

\author{Sarah Goler}
\affiliation{NEST Istituto Nanoscienze CNR and Scuola Normale Superiore, I-56127, Pisa Italy}

\author{Trevor D. Rhone}
\affiliation{Department of Physics and Department of Applied Physics
and Applied Mathematics, Columbia University New York, USA}

\author{Melinda Han}
\affiliation{Department of Physics and Department of Applied Physics
and Applied Mathematics, Columbia University New York, USA}

\author{Rui He}
\affiliation{Department of Physics and Department of Applied Physics
and Applied Mathematics, Columbia University New York, USA}

\author{Philip Kim}
\affiliation{Department of Physics and Department of Applied Physics
and Applied Mathematics, Columbia University New York, USA}

\author{Vittorio Pellegrini}
\affiliation{NEST Istituto Nanoscienze CNR and Scuola Normale Superiore, I-56127, Pisa Italy}

\author{Aron Pinczuk}
\affiliation{Department of Physics and Department of Applied Physics
and Applied Mathematics, Columbia University New York, USA}

\begin{abstract}
{Coherent coupling of Dirac fermion magneto-excitons with an optical phonon is observed in graphite
as marked magnetic-field dependent splittings and anti-crossing behavior of the two coupled modes. The sharp magneto-phonon resonance occurs in regions of the graphite sample with properties of superior single-layer graphene having enhanced lifetimes of Dirac fermions. The greatly reduced carrier broadening to values below the graphene electron-phonon coupling constant explains the appearance of sharp resonances that reveal a fundamental interaction of Dirac fermions.}
\end{abstract}
\maketitle
\bigskip
\bigskip

The quest for ultra-high quality graphene,  the new celebrated
two-dimensional (2D) electron gas \cite{geim}, is driven by
expectations of discoveries of novel physics and applications linked
to massless Dirac fermions \cite{andrei1,kim}. A class of quantum
coherent effects dictated to the rather weak electron-phonon interaction \cite{jun0,pisana} is linked 
to the coherent coupling of the particle-hole transitions
with optical phonons resonantly tuned by an external
magnetic field \cite{ando,falko}. 
\par
These resonant coupling phenomena 
(magneto-phonon resonances or MPR)
occur when the energy spacing between Landau
levels (LLs) is continuously tuned to cross the energy of an optical
phonon mode. MPRs have been largely explored in bulk semiconductor
materials \cite{leadley, barnes, johnson}, in two-dimensional
semiconductor systems and in quantum dots \cite{chang,hameau}.
\par
Recent theoretical studies in graphene have suggested that MPR leads to a 
rich splitting and anti-crossing phenomena 
of the even parity $E_{2g}$ long wavelength optical mode due to
its MPR interactions with excitations across Dirac fermion Landau
levels (magneto-excitons) \cite{ando,falko}.
\par
The experimental manifestation of the quantum interactions in
the MPR relies on a delicate interplay between the lifetimes of Dirac fermion modes and the
electron-phonon coupling. A magnetic field tunable anomaly in the optical phonon
of graphene has been reported \cite{yan, faugeras}. In these works the
resolution of the anti-crossing that is the signature of the coherent coupling of electron-phonon modes is impeded by the relatively short lifetimes of the fermion excitations. The manifestation of the sharp MPR requires damping of the Dirac fermion excitations well below the weak electron-phonon coupling. This qualitatively new regime of interacting Dirac Fermion modes has not been achieved so far.
\par
In this letter we demonstrate the coherent mixing of the $E_{2g}$ phonon and the Dirac fermion magneto-exciton mode 
that is the essence of the MPR. At selected values of the magnetic field we observe the emergence of two distinct modes displaying the characteristic anti-crossing behavior. These observations uncover unprecedented narrow linewidths of Dirac fermion magneto-exciton transitions. The Landau level transition Lorentzian widths of 3.2 meV reported here are about three times smaller than those of previous investigations at similar magnetic fields \cite{faugeras,andrei,orlita} and yield damping values below the graphene electron-phonon coupling constant. 
Surprisingly, the enhanced Dirac fermion lifetimes required for the observations of
the characteristic MPR electron-phonon mode anti-crossings are found in selected regions
of graphite.
\par
We show below that our findings can be modeled with a simple coupled
mode Hamiltonian describing the $E_{2g}$ mode and sharp magneto-excitons of a
graphene layer. The sharp Landau level transitions found in this
work and the interpretation in terms of a single-layer graphene
model are in line with recent experimental and theoretical studies
that suggest the existence of high-quality decoupled graphene flakes in
bulk graphite \cite{andrei,potemski,lopes}.
In our work the unique areas are identified
by the capability to scan regions of the sample within the environment of the
magneto-optics experiment. Consistent with this
interpretation, our spatially-resolved Raman analysis reveals a large
non-uniformity of the graphite sample and areas of the sample where no
MPRs are identified (see Supplementary).
\par
The main MPR is found close to 5T and involves the
${-1}\rightarrow {2}$ and ${-2}\rightarrow {1}$ Landau level transitions of
Dirac fermions (Landau levels in the $\pi^*$ ($\pi$) conduction
(valence) band correspond to the positive (negative) integers) as shown in Fig.1a.
Representative examples of spectra showing the two coupled modes
$\omega _- $ and $\omega _+$ are presented in Fig.1c and d
together with the magnetic-field evolution of the expected mode
anti-crossing behavior (Fig.1 b). The simple coupled mode physics
underlying the data enables algebraic extraction of parameters of
the interacting system including the Dirac fermion lifetimes.

\par
\begin{figure}
\begin{center}
\includegraphics*[width=7.5cm]{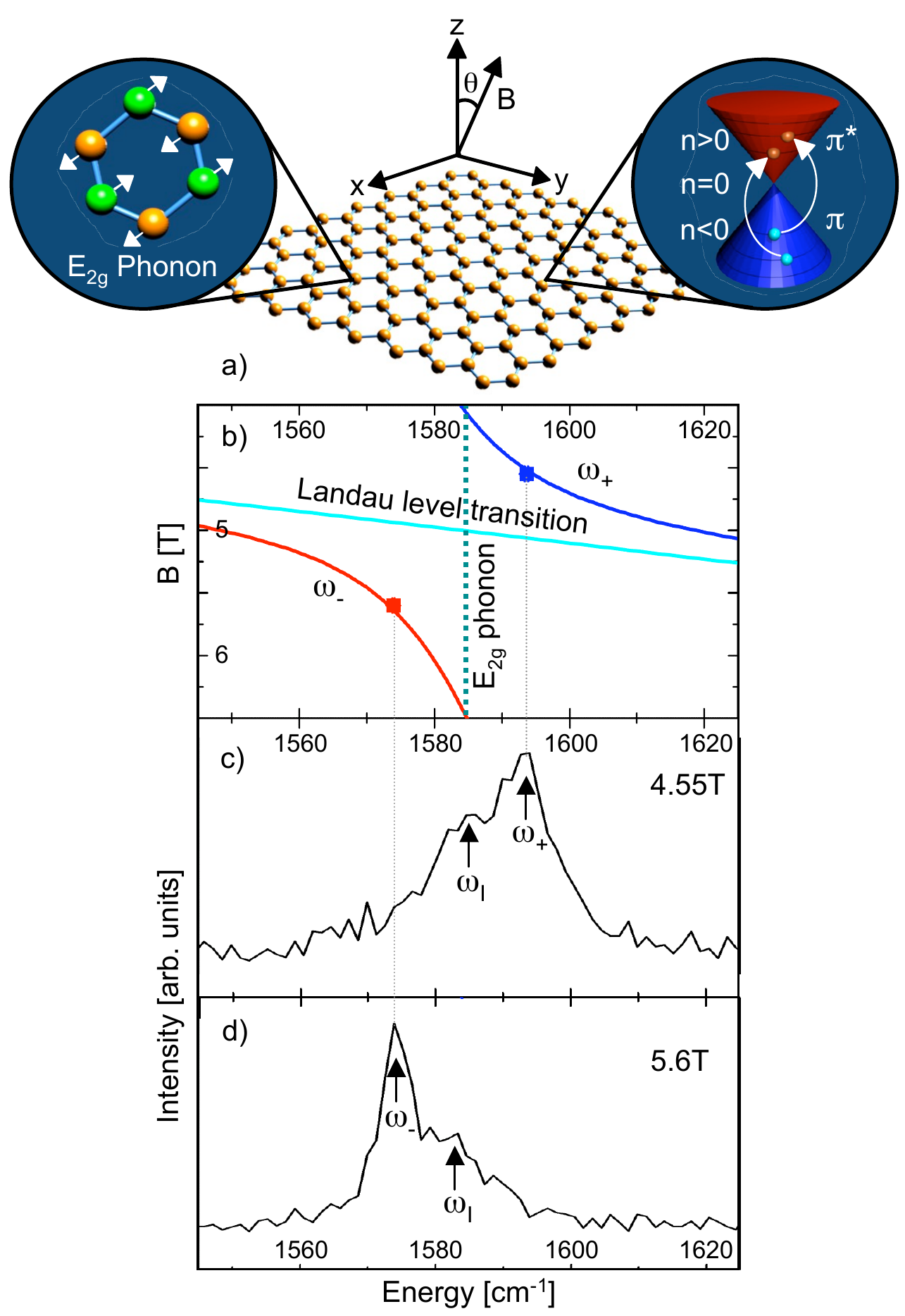}
\end{center}
\caption{ 
{\bf a} Schematic visualizations of the phonon and magneto-exciton modes involved in the magneto-phonon resonance (MPR). The graphene layer is illustrated in the x-y plane.
The magnetic-field direction in the experiment is tilted by $\theta = 20^{\circ }$ from the z-axis.
{\bf b} Energies of two coupled normal modes versus magnetic field close to the MPR
at around  B = 5T. {\bf c} and {\bf d} Raman spectra acquired at 4.55T and 5.6T, respectively.  The energies of the two
coupled modes $\omega _-$ and $\omega _+$ are indicated by the red and blue dots in panel b.  The origin of the
additional peak $\omega _I$ is discussed later in the text.}
\end{figure}
\par
Raman experiments were performed at 2K in a magnetic field up to 14T
in a backscattering configuration with a tilt angle of $\theta $ =
20$^\circ$ between the magnetic field and the c-axis direction of
Kish graphite (Toshiba ceramics) which was deposited onto a $Si/SiO_2$ substrate (see
Fig.1a). A diode-pumped solid-state laser with a 532nm emission
line was used as the excitation source and focussed on the sample
with a spot size of about $80\mu m$ and a power of around $30 mW$.
The laser spot was moved around on the sample to select regions with
optimized MPR responses. We also performed micro-Raman measurements
(spatial resolution of $0.5 \mu m$) at room temperature in air on a
different piece of the sample which exhibits large spatial
variations of the 2D band (see Supplementary).
\par
Figure 2 displays the magnetic-field dependence of the Raman spectra
that we obtained on two different spots of the sample.
The $E_{2g}$ phonon displays drastically different behavior. In
Fig.2a the spectra are very sensitive to the tuning of the magnetic
field, while in Fig.2b no changes are observed. This
highlights the non-uniformity of the graphite and the presence
of regions that behave like single-layer graphene. Further support
for this picture comes from the spatially-dependent analysis of the
Raman spectra that is reported in the supplementary material
section. Behavior similar to that in Fig.2b has been reported and
ascribed to the usual Bernal stacked graphite \cite{faugeras}.
Below we will focus on the remarkable data displayed in
Fig.2a.
\par
The magnetic-field evolution of the mode energies of these spectra are shown in Fig.2c.
Previous studies have observed a phonon energy modulation of about 40
cm$^{-1}$ using magnetic fields up to 30 Tesla \cite{faugeras}. Here
we found the optical phonon energy is modulated by more than 60
cm$^{-1}$ with fields less than 6 Tesla, indicating the presence of
very high quality Dirac fermions with long lifetimes in the sample.
\par
\begin{figure}
\begin{center}
\includegraphics*[width=7.5cm]{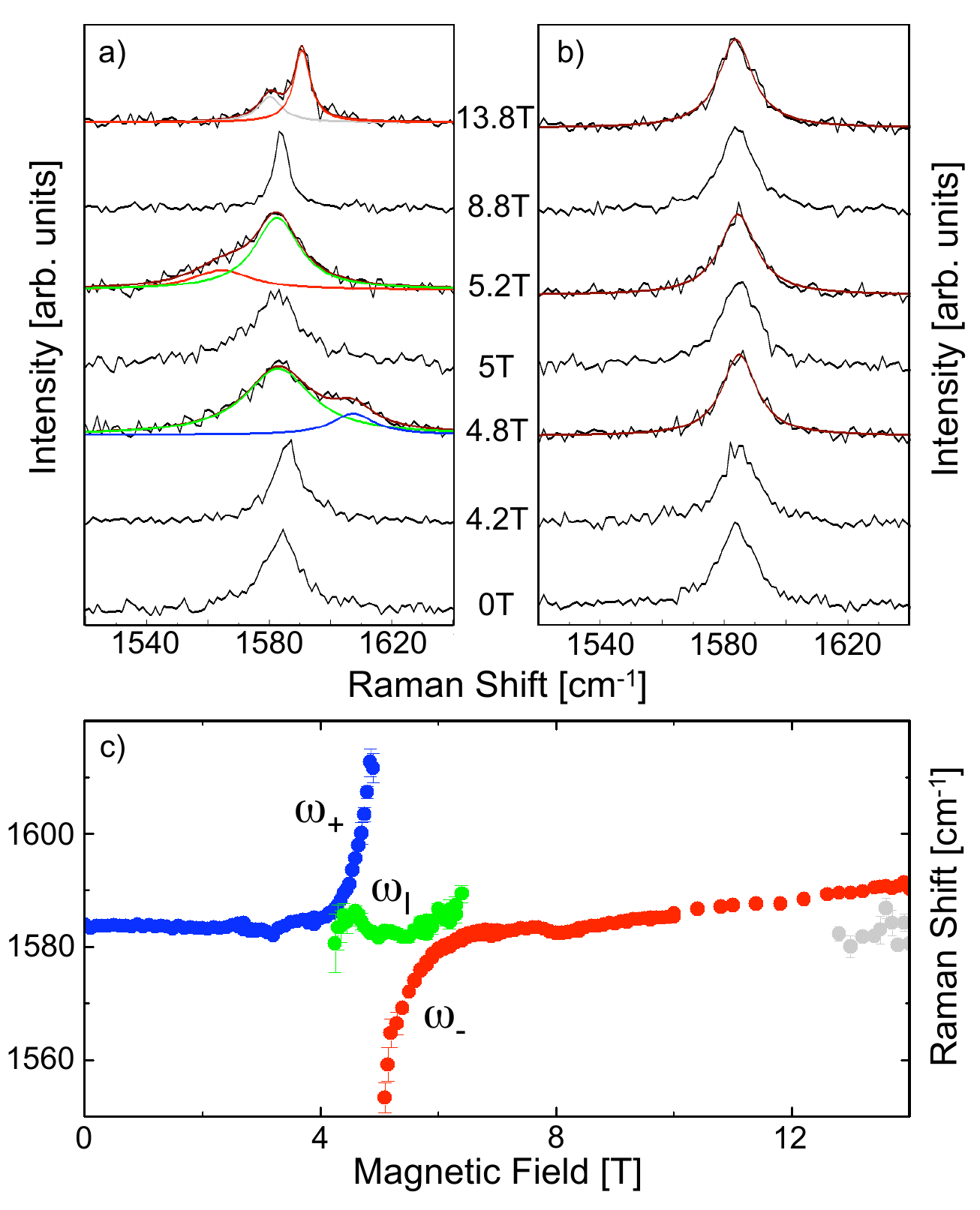}
\end{center}
\caption{{\bf a} and {\bf b}. Magnetic field dependence of the $E_{2g}$ phonon spectra taken at two representative 
regions of the sample. One region displays significant MPR effect while the other doesn't. 
The smooth curves are Lorentzian fits to the data. {\bf c}. Mode energy evolution extracted from the spectra 
displayed in {\bf a}. The colors of the four branches correspond to those used in the Lorentzian fits.}
\end{figure}
\par
The extreme sensitivity of the spectral lineshape to magnetic field
changes seen in Fig.2a suggests the existence of LL transitions
that come into and out of resonance with the phonon energy. It is
also remarkable that at $\sim$6.5 Tesla the phonon becomes extremely
sharp, with a width of 4cm$^{-1}$. To the best of our knowledge this is the
narrowest $E_{2g}$ phonon width reported in graphene-related
materials.
\par
Electric-field effect studies have shown that a sharp $E_{2g}$
phonon is linked to the absence of resonant electron-hole pair
transition states that the lattice vibration can decay into
\cite{jun0,pisana,ando1}. This indicates that at around 6.5 Tesla
no LL transition exists that resonantly interacts with the
phonon. It follows that observations of a very sharp phonon have a
nontrivial implication on the LL structure of the underlying
electronic system. Furthermore, the significant lineshape changes
around 5T suggest that a LL transition with large degeneracy comes
into resonance with the long wavelength optical phonon. Comparing with
current LL studies of graphitic materials
\cite{toy,orlita2,orlita3}, the transition matches well with the
${-1}\rightarrow {2}$ (and ${-2}\rightarrow {1}$) transitions of
Dirac fermions. Based on these considerations we proceed to model
the electron system in the MPR anticrossing displayed in Figs. 2a
and 2c as Dirac fermions with discrete LL structures charateristic
of those in single-layer graphene of high perfection.
\par
Figure 3 shows the detailed evolution of the phonon spectrum. We
observe that the lineshape is very asymmetric and the asymmetry
changes side when crossing 5 Tesla. As shown, all the asymmetric
spectra can be decomposed into two Lorentzian peaks.
\begin{figure}
\begin{center}
\includegraphics*[width=7.5cm]{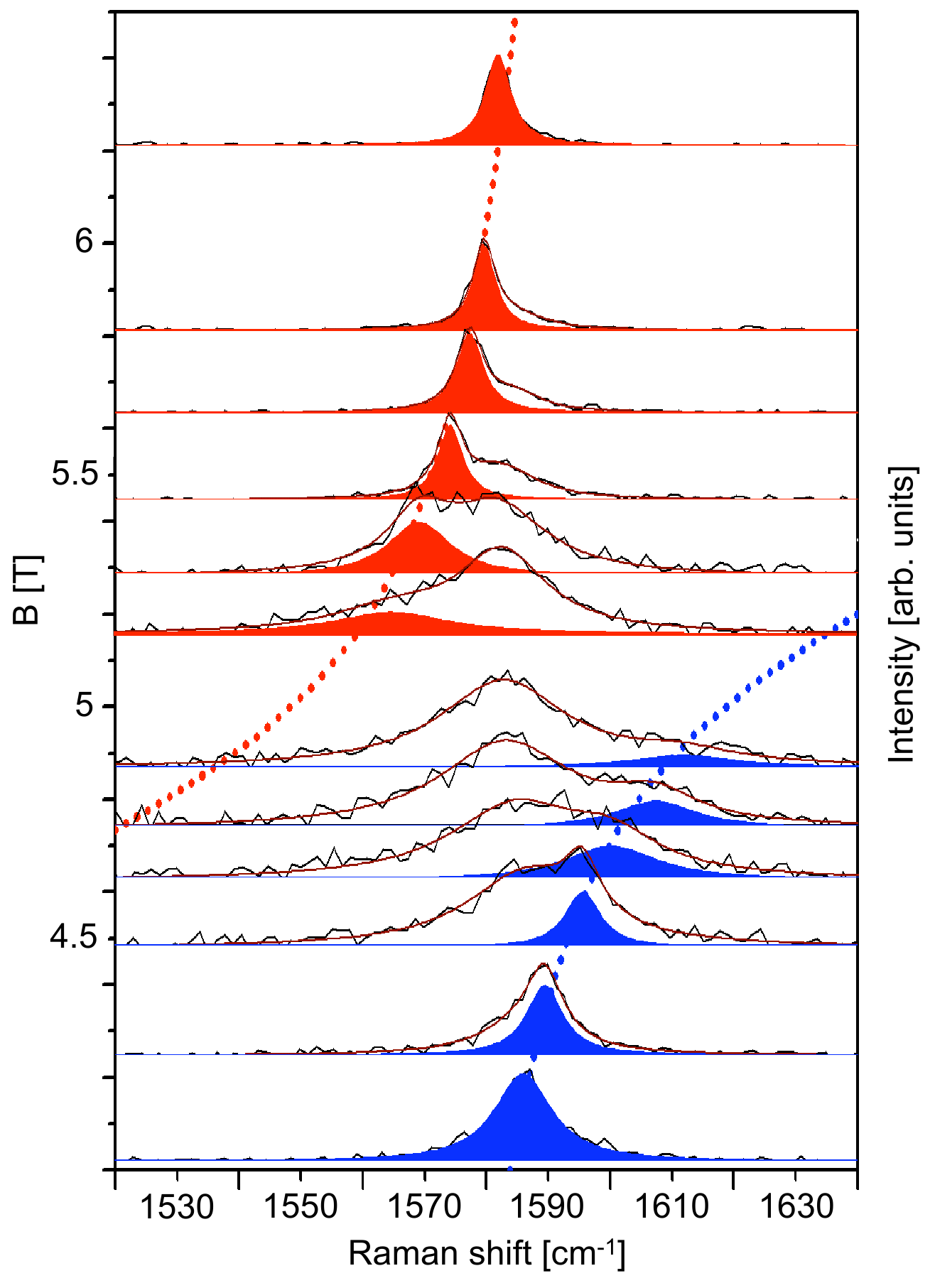}
\end{center}
\caption{Detailed Raman spectra of the main MPR observed for magnetic fields ranging from 4.2T to 6.4T. The
theoretical calculation of the anti-crossing associated to the magneto-phonon resonance (MPR) at 5T (see text) is
indicated by red and blue dotted lines. The spectral component that corresponds to the two
coupled modes is highlighted.  The spectra are shifted vertically so that each red and blue
component taken at a given magnetic field has its peak aligned at that magnetic field.}
\end{figure}
The red and blue modes which we highlight in the figure form a
paired behavior at about 5 Tesla. Taking into account the
20$^{\circ}$ angle between the magnetic field and the graphite
c-axis, at resonance the field perpendicular to the basal plane is
4.7 Tesla. The resonance condition of the ${-1}\rightarrow {2}$ LL
transition of Dirac fermions gives $(1+\sqrt{2})\frac{\sqrt{2}\hbar
v_F}{l_B}=196$ meV where $l_B$ is the magnetic length, 196 meV is
the $E_{2g}$ phonon energy. This determines the Fermi velocity $v_F$
to be 1.03$\times10^6$m/s, in good agreement with studies of Dirac
fermions in graphitic materials.
\par
The mode energy and linewidth obtained from the analysis
are displayed in Fig.4. The splitting of the two anti-crossing branches
asymptotically approaches a large value of 70 cm$^{-1}$ at the
resonance field of 5 Tesla. This is in contrast with results in Fig.2b 
and the studies reported in Ref.\cite{faugeras}, in which the impact
of the LL transition is seen largely as a weak perturbation that
only weakly renormalizes the phonon energy and broadens the phonon
Raman peak. In the results shown in Fig.4, the sharpness of the LL
transition leads to the strong mixing of the optical phonon with a
magneto-exciton, and to the establishment of two distinct new
collective modes that have characteristic anti-crossing behavior.
\begin{figure}
\begin{center}
\includegraphics*[width=7.5cm]{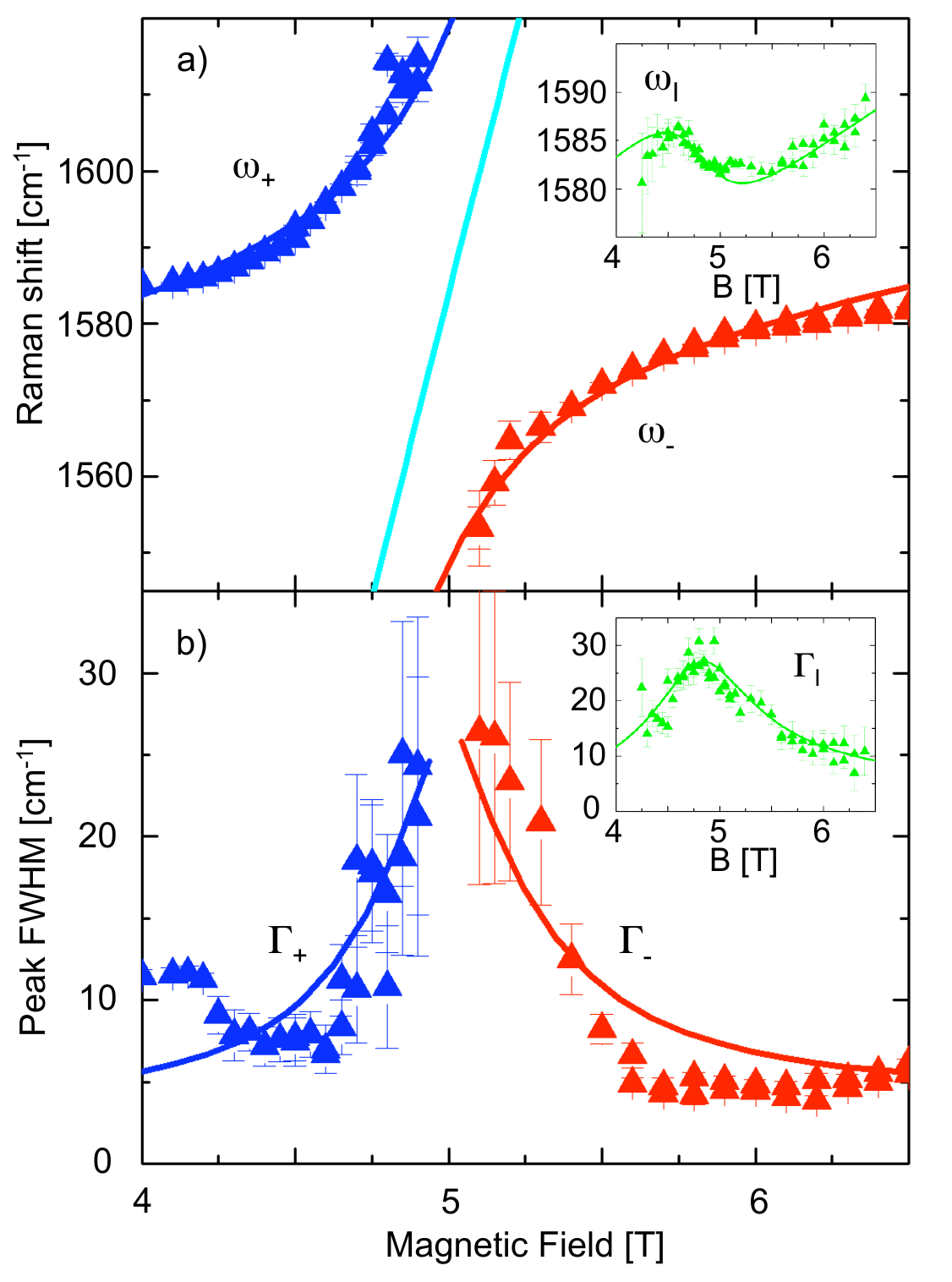}
\end{center}
\caption{MPR induced anti-crossing behavior of the coupled modes. 
{\bf a} and {\bf b} show the mode energy and full-width-half-maximum (FWHM), respectively.
The blue and red lines are the theoretical fits of the energy of
the two  branches.  The cyan line in {\bf a} is the Landau level transition ${-1}\rightarrow {2}$ or ${-2}\rightarrow {1}$.  The
blue and red triangles are the experimental data.  The inset reports the magnetic-field evolution of the $\omega _I$ mode.}
\end{figure}
\par
To quantitatively analyze our data, we employ a coupled mode theory
that was proposed by Goerbig et al. in Ref.\cite{falko} where the
two coupled normal modes are described as
\begin{eqnarray}
\centering
\hbar \omega_{\pm}=\frac{E_{\text{PH}}+E_{\text{ME}}}{2}\pm\sqrt{(\frac{E_{\text{PH}}-E_{\text{ME}}}{2})^2+g^2}.
 \label{eq1}
\end{eqnarray}
In the equation, $E_{\text{PH}}$ and $E_{\text{ME}}$ describe the
phonon and magneto-exciton respectively, while $g$ is the coupling
parameter. To explicitly describe the broadening of the modes, we
let $E_{\text{PH}}=\omega-i\gamma$, $E_{\text{ME}}=\Omega-i\Gamma$
where $\omega$, $\Omega$, $\gamma$ and $\Gamma$ are energies and
half widths of the phonon and magneto-exciton.
\par
At resonance $\omega=\Omega$, $\hbar
\omega^{r}_{\pm}=\omega\pm\sqrt{g^2-(\frac{\Gamma-\gamma}{2})^2}-i\frac{\Gamma+\gamma}{2}$.
The phonon width $\gamma$ in the absence of Landau damping into
electron-hole pairs is 4cm$^{-1}$ so $\gamma=2 cm^{-1}$. The
analysis using data in Fig.4 yields the energy and broadening of the two coupled modes
at resonance: $\hbar \omega^{r}_{+}=(1619-i14) cm^{-1}$ and $\hbar
\omega^{r}_{-}=(1549-i14) cm^{-1}$. These experimental data allow us
to algebraically determine all the parameters in the coupled mode
Hamiltonian: $\omega=1584 cm^{-1}$, $\Gamma=26 cm^{-1}$ and $g=37
cm^{-1}$.
\par
The value of $g$ provides a direct measurement of the
electron-phonon coupling strength since it is directly linked to the
dimensionless electron-phonon coupling constant $\lambda$
\cite{ando} by $g=\sqrt{\frac{\lambda}{2}}\hbar\omega_B$ where
$\hbar\omega_B=\frac{\sqrt{2}\hbar v_F}{l_B}$. Using the
experimental result $(\sqrt2+1)\hbar\omega_B=\frac{1619+1549}{2}
cm^{-1}$, we find $\lambda=6.36\times10^{-3}$. Physically, this
coupling strength reflects the modulation rate of the nearest
neighbor hopping integral $\gamma_0$ with respect to the stretching
of the carbon-carbon bond length $b$ and we obtain
$\frac{d\gamma_0}{db}=6.66$eV/\AA. These values are in reasonable
agreement with experiments on monolayer graphene \cite{yanssc} and
epitaxial graphene\cite{faugeras}.
\par
The Landau level transition half width of 26cm$^{-1}$ (roughly 3
meV) is an important indication that the Dirac fermions residing in
graphite are of a very high degree of perfection. As a
comparison, in epitaxial graphene with a reported mobility of
250,000cm$^2$/(V$\cdot$s) \cite{orlita}, the Landau level width at
5 Tesla is about 10meV. 
\par
The narrow width $\Gamma $ less than $g$, a
condition not met before, allows for the coherent coupling between
the phonon and the fermion excitations leading to the anti-crossing
phenomenon seen in our data. The observation of the delicate MPR
coherent processes in graphite opens further venues for
fundamental optics research on Dirac fermions.
\par
We now briefly discuss the green mode labelled $\omega _I$ with
energy and width displayed in the insets to Fig.4. The behavior of
this mode is quite similar to the magnetic oscillations observed in
Ref. \cite{faugeras}. Interestingly, we found that the resonance
field of $\omega _I$ is slightly lower (at 4.8T) than that of 
the two anti-crossing modes (at 5T). This might indicate disorder-induced
Fermi velocity renormalization of graphene. The existence of the the
$\omega_I$ mode is a further indication of sample non-uniformity.
\par
For the magnetic field range below 4 Tesla, magneto-phonon resonance
with interband transitions of higher LL indexes are expected.
However, we were unable to clearly identify such resonances,
possibly due to the smaller LL degeneracy as well as limitations
imposed by our signal to noise ratio detection limit.
\par
Starting at about 13 Tesla, the phonon splits again indicating the
occurrence of another anti-crossing (Fig.2c). This resonance could be
due to the $0\rightarrow2$ LL transition that is expected at 14.6
Tesla or the $0\rightarrow1$ transition expected at 29.1 Tesla.
Because the resonance is incomplete, we are unable to rule out one
or the other unambiguously.

\par 

{\bf Supplementary materials}

\par

{\bf Spatially-resolved Raman analysis of the 2D phonon peak in Kish graphite}

\par
Kish graphite was deposited on a $Si/SiO_2$ substrate using the standard micromechanical cleavage method.  Spatially-resolved Raman spectroscopy measurements were acquired in ambient conditions.  An argon ion laser emitting at 488 nm, used to excite the phonons in the graphite sample, was focused to a spot size of about 0.5 $\mu m$ with a power of about 20 mW.  The position of the sample was controlled by piezoelectric stages.  Visual feedback was provided by a CCD camera, allowing for precise control of the position and focus of the laser on the sample.  Raman measurements were obtained every 1 $\mu m$, over an area of 40 $\mu m ^2$, with an acquisition time of one minute.
\par
\par
\begin{figure}
\begin{center}
\includegraphics[width=9cm]{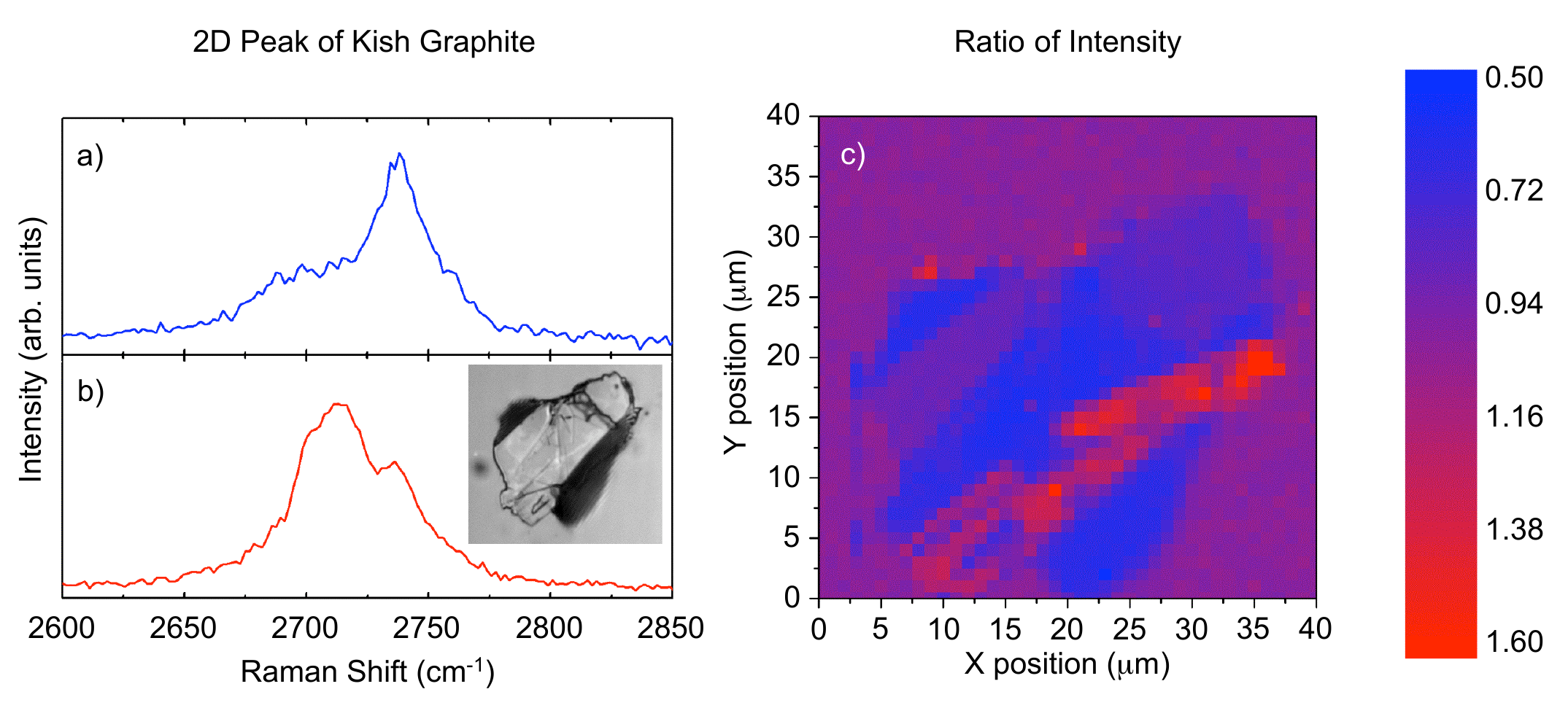}
\end{center}
\caption{a)  The characteristic 2D peak of graphite.
b)  Another 2D peak obtained on the same piece of Kish graphite shown in the inset.
c)  A map of the ratio of intensities of the double peak structure of the 2D Kish graphite peak was obtained by integrating over energy ranges centered on the two components of the 2D peak.  The lower energy component was integrated from 2704 $cm^{-1}$ to 2718 $cm^{-1}$ or 9 pixels of the CCD camera.  An equivalent number of pixels were selected in the high energy range, 2727 $cm^{-1}$ to 2742 $cm^{-1}$.  The spatially-resolved Raman spectra were acquired with a step size of 1 $\mu m$.  The blue regions correspond to peaks resembling panel a and the red areas or similar to panel b.}
\end{figure}
\par
We focus our spatially-resolved investigations on the 2D peak. In fact, the 2D peak of graphite is extremely sensitive to the number of layers present in the sample \cite{ferrari, ferrarissc}.  It is one of the trademark Raman signatures used to distinguish between single and bilayer graphene \cite{ferrari, ferrarissc}.  In monolayer graphene, the 2D peak is a single symmetric peak whereas in bilayer graphene the 2D peak can be fitted with four peaks, evidence of the four possible double resonance processes that occur due to the splitting of the valence and conduction bands in the vicinity of the Dirac points \cite{ferrari, ferrarissc, reich}.  Analysis of the 2D peak of our sample exhibits variations from the standard graphite peak shown in Fig.5a.  The double peak structure shifts from including the more intense peak on the higher energy side to the lower energy side.  The lower energy peak component has been shown by LukÕyanchuk, et al., to correspond to the single layer 2D peak position \cite{luk}.  The nonuniformity of the 2D peak is a strong indication of the decoupling of the layers leading to areas of the sample with a high contribution of electronically decoupled layers.
\par
The ratio between the intensity of the peak positions of the two components comprising the 2D peak of graphite are plotted as a function of position in Fig.5c.  The regions in which the 2D peak corresponds to the characteristic graphite 2D spectrum (Fig.5a) are indicated by the blue areas in Fig.5c.  The spectra in which the lower energy component dominates (Fig.5b) are red in Fig.5c.  Variations of the intensity of the two peaks is a clear indication of the irregular single layer contribution to the Raman signal.

{\bf Acknowledgement } The authors would like to thank V. Piazza
for assistance in the spatially-resolved measurements. This work is
supported by ONR (N000140610138 and Graphene MURI), by NSF
(CHE-0641523), and the NYSTAR. P.K. acknowledges support from the
FENA MARCO Center. A. P. is supported by NSF (DMR-0803691), and a grant
from the Keck Foundation. V.P is supported by a grant of the Ministry of Foreign Affairs (Italy).
%\end{addendum}

\end{document}